\documentclass[11pt]{article}
\usepackage[margin=1in]{geometry}
\usepackage{graphicx}
\usepackage{siunitx}
\usepackage{chemformula}
\usepackage{ulem}
\usepackage{url}
\usepackage{float}
\usepackage{hhline}
\usepackage[nomargin,inline,marginclue,draft]{fixme}
\graphicspath{ {images/} }

\def\Title#1{\begin{center} {\Large {\bf #1} } \end{center}}
\def\Author#1{\begin{center} {\normalsize {\sc #1} } \end{center}}
\def\Institution#1{\begin{center} {\normalsize {\it #1} } \end{center}}
\def\Abstract#1{\noindent {\normalsize {\bf Abstract:} {\normalfont #1}}}
\def\Conference{\vspace{4mm}\begin{raggedright} {\normalsize {\it Poster presented at the 2019 Meeting of the Division of Particles and Fields of the American Physical Society (DPF2019), July 29--August 2, 2019, Northeastern University, Boston, C1907293.} } \end{raggedright}\vspace{4mm}}

\begin{document}

%
%

\Title{Simulation of Scintillator Counters\\with Embedded Wavelength-Shifting Fibers}

\Author{Ralf Ehrlich}

\Institution{University of Virginia, Department of Physics, Charlottesville, Virginia 22904}

\Abstract{We describe a complete end-to-end simulation of the response of scintillator counters to charged particles. The counters were extruded with a titanium dioxide surface coating and two channels for the embedded wavelength-shifting fibers which are read out by silicon photomultipliers. The simulation includes the production and propagation of scintillation and Cerenkov photons, the response of the silicon photomultipliers, and the generation of the signal waveforms. Probability lookup tables are used to speed up the simulation of the photon propagation inside the counters. The simulation was tuned to match measured data obtained from a test-beam study in the Fermilab Meson Test Beam Facility using 120 GeV protons. The counters are intended to be used in the cosmic-ray veto detector for the Mu2e experiment at Fermilab.}

\Conference

%
%
\section{The Mu2e Experiment}\label{ch:TheMu2eExperiment}

The Mu2e experiment will search for coherent muon-to-electron conversions in the orbit of aluminum nuclei~\cite{Mu2eTDR}. The monoenergetic electrons from such conversions have an energy just below the muon mass. A major source of background comes from cosmic rays, which can produce electrons, positrons, or muons that mimick conversion electrons. In order to remove these backgrounds, a cosmic ray veto system (CRV) will be implemented. The CRV consists of an active detector which surrounds the Mu2e detector. This CRV will be used to veto events that look like conversion electrons and coincide with a cosmic ray muon passing through the CRV. The goal is to achieve a veto efficiency of 99.99\%. 

The CRV is made of 5,376 individual CRV counters. An accurate computer model of how these CRV counters respond to charged particles is essential for the design of the CRV. This includes determining the veto efficiency of the CRV, and where coverage is needed. Simulations are used to estimate the number of cosmic-ray muons which cannot be vetoed by the CRV, and to estimate the dead-time due to the large number of false positives caused by the radiation background.

\section{CRV Counters}\label{ch:CRVCounters}

\begin{figure}
	\centering
	\includegraphics[width=0.65\textwidth,keepaspectratio]{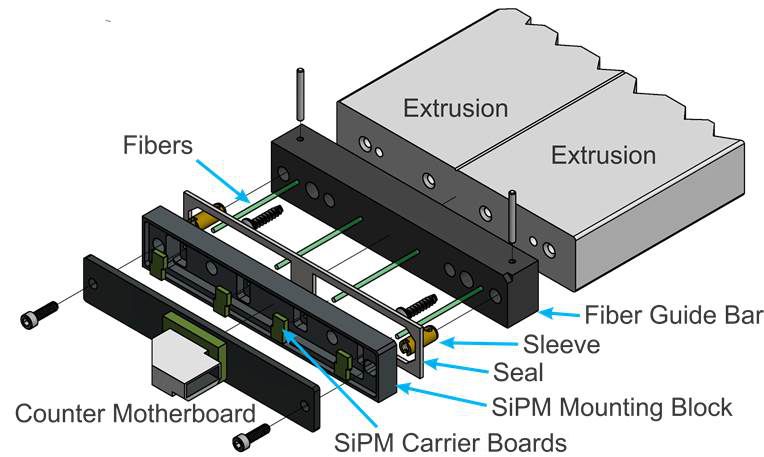}
	\caption{Exploded view of one of the two readout ends of a CRV di-counter.}
	\label{fig:dicounter}
\end{figure}

The cosmic ray veto counters are \SI{51}{mm} wide, \SI{20}{mm} thick polystyrene scintillators with two holes for the wavelength-shifting fibers (Fig.~\ref{fig:dicounter}). The embedded wavelength-shifting fibers have a diameter of \SI{1.4}{mm}. The counters have lengths between \SI{0.9}{m} and \SI{6.9}{m}. Two counters are glued side-by-side to form a di-counter. The di-counter is terminated at both ends by a fiber guide bar with holes into which the fibers are glued. The surfaces of the fiber ends are fly cut. A readout manifold with four silicon photomultipliers (SiPMs) and a small electronics board is mounted on the fiber guide bars.~\cite{Testbeam2016}. Some CRV di-counters have a mirror manifold
(with a reflector at the fiber ends) instead of the readout manifold at one end either because of limited space or because the di-counter end is located in a region with high radiation levels.

The scintillator used for the CRV is made of polystyrene doped with 1\% PPO and 0.03\% POPOP~\cite{Testbeam2016}. The extrusions were produced at the Fermilab-NICADD Extrusion Line Facility~\cite{Scintillator}. The extrusions are co-extruded with a \SI{0.25}{mm} thick reflective layer containing 30\% \ch{TiO2} and 70\% polystyrene~\cite{Testbeam2016}. The wavelength-shifting fibers are double cladded non-S type fibers made by Kuraray~\cite{Kuraray} with a diameter of \SI{1.4}{mm} (Table~\ref{table:WLSfiber}). The dopant is \SI{175}{ppm} Y11. 

The SiPMs are Hamamatsu S13360-2050VE, which are \SI{2}{mm} x \SI{2}{mm} multipixel photon counters with 1584 pixels each \SI{50}{{\mu}m^2}~\cite{HamamatsuS13360}. 

\begin{table}
	\begin{tabular}{l l l l}
		\hhline{====}
		& Fiber core & Inner cladding & Outer cladding \\ \hline
		Material & Polystyrene & PMMA & Fluorinated polymer \\
		Index of refraction at \SI{589}{nm} & 1.59 & 1.49 & 1.42 \\
		Radius& \SI{0.616}{mm} & \SI{0.658}{mm} & \SI{0.700}{mm} \\ \hhline{====}
	\end{tabular}
	\caption{Properties of the wavelength shifting fibers~\cite{Kuraray}}
	\label{table:WLSfiber}
\end{table}

\section{Test beam}\label{ch:Testbeam}

Test beams that are suitable for all different counter lengths were not available. Hence, only 3-m-long counters were tested, and simulations were used to determine the performance of counters of different lengths. Hence, it is vital that the simulations of the CRV response to cosmic-ray muons be as accurate as possible. Every effort was made to validate the scintillator response using real data.

The results of the simulations discussed in this paper are compared with test-beam data acquired at the MTest beamline in June 2017 at Fermilab (Ref.~\cite{TestbeamReport}). The setup of this test beam is similar to the one discussed in Ref.~\cite{Testbeam2016}. A \SI{120}{GeV} proton beam was directed at specific locations of the CRV counter, and the counter response was recorded using the same front-end electronics as will be used in the experiment. The counter width was \SI{49}{mm}, slightly less than the \SI{51}{mm} production counter width.

\section{CRV Counter Simulation}\label{ch:CRVSimulationChain}

The CRV counter simulation happens in the following sequence:

\begin{enumerate}
	\item Particles are transported through a CRV counter by GEANT4~\cite{GEANT4}, which takes care of all particle interactions, decays, energy depositions, etc. GEANT4's scintillation and Cerenkov processes are not used, which means that GEANT4 is not being used to simulate photons. They are handled separately in the next step.
	\item The positions, times, momenta, and deposited energies of the primary and all secondary particles are used to generate scintillation photons and Cerenkov photons along the particles' trajectories. Lookup tables are used to determine the probabilities for these photons to be detected by one of the four SiPMs of a CRV counter. The arrival times of these photons at the SiPMs are also calculated using lookup tables. The simulation of these photons was done with lookup tables instead of GEANT4, since a full GEANT4 simulation of every photon is too time consuming. More details about these lookup tables are given in the next section.
	\item The photons arriving at the SiPMs are distributed to individual pixels according to a pixel distribution map (Fig.~\ref{fig:photonMap}), which was found by a GEANT4 simulation of the photon distribution at the fiber ends. The SiPM simulation includes saturation effects and the pixel recharge behavior, dark counts, cross-talk, and after-pulses. The results of this simulation step are the times and released charges of the avalanches of individual pixels. 
	\begin{figure}[H]
		\centering
		\includegraphics[width=0.5\textwidth,keepaspectratio]{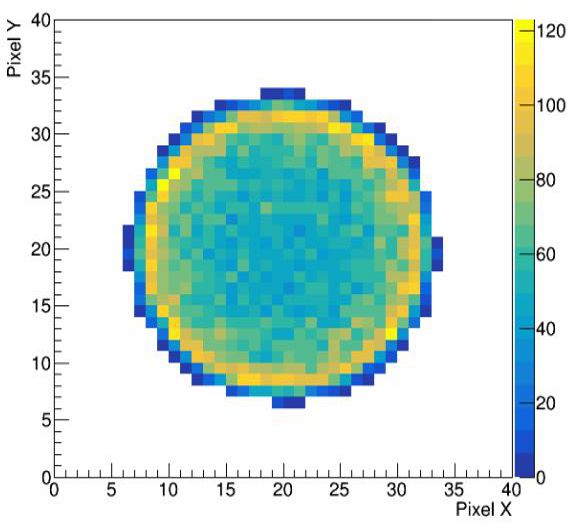}
		\caption{Simulated photon distribution at the Hamamatsu SiPM S13360-2050VE.}
		\label{fig:photonMap}
	\end{figure}	
	\item The front-end electronics response to the SiPM is determined. This is a superposition of single-pixel waveforms occurring for every avalanche at individual pixels of the SiPM. The value of the resulting waveform is digitized and recorded every \SI{12.55}{ns} (Fig.~\ref{fig:SampleEvent}).
	\begin{figure}[H]
		\centering
		\includegraphics[width=0.7\textwidth,keepaspectratio]{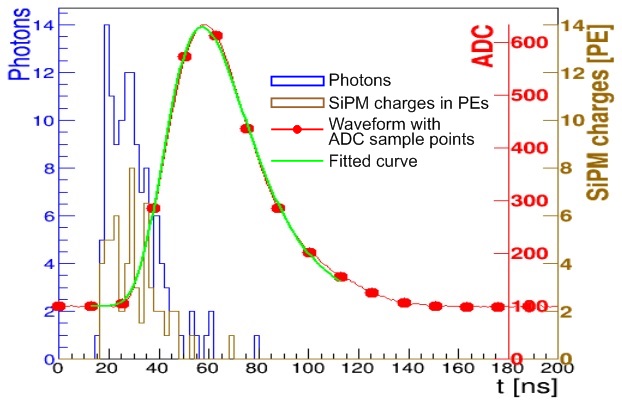}
		\caption{Simulation result of the response of the front-end electronics to an \SI{120}{GeV} proton normally incident at the center between both fibers and \SI{2}{m} away from the SiPM. Blue: photons which were detected by the SiPM; yellow: output charges of the SiPM in number of PEs; red: ADC waveform of the SiPM with the points which were recorded every \SI{12.55}{ns}; green: fitted curve to the points.}
		\label{fig:SampleEvent}
	\end{figure}
\end{enumerate}

The output of the last simulation step is similar to the output of the real front-end electronics. As is done with real data, a curve is fitted to the digitized samples of each SiPM in order to estimate the pulse times and the number of photo electrons which caused this pulse. 

\section{Lookup Tables}\label{ch:LookupTables}

A full GEANT4 simulation of particles passing through one CRV counter, including all photons being produced in and traveling through the scintillator and fibers, takes several minutes and hence was not feasible as our goal was to simulate the muon flux for the full detector over several experiment run times. Therefore, another approach was taken: lookup tables.

The CRV counter is divided into a 3D grid, and every bin of the grid contains the following information for the scintillation photons created within this bin: 
\begin{itemize}
	\itemsep0em
	\item the probability of these photons to get detected by a SiPM,
	\item the photon travel time distribution, and
	\item the distribution of the number of wavelength-shifts in the wavelength-shifting fiber (which increases the total travel time of the photons). 
\end{itemize}
When a charged particle passes through a bin of the CRV counter, the energy that this particle deposits in this bin (provided by GEANT4) is used to calculated the number of scintillator photons created in this bin. The detected number of photons at every SiPM and the detection times are determined based on the probability information stored in the lookup table for this bin. The position of the bin in the 3D grid serves as lookup parameter for the lookup table. With this method, GEANT4 does not have to produce and trace individual photons inside of the CRV counter, which speeds up the simulation significantly. Different lookup tables are produced for the 9 different counter lengths, with and without reflectors.

The lookup tables are built using a standalone GEANT4 simulation which models the entire CRV counter with all of its material properties. If material parameters could not be found, they had to be tuned in such a way that the simulation results agree with data from test-beam measurements. GEANT4 randomly generates 500,000 scintillation photons in each bin of the 3D grid with a uniformly distributed position, direction, and polarization, and an energy spectrum based on the scintillation spectrum. The bin widths vary between \SI{0.5}{mm} (e.g. close to the fiber holes) and \SI{3.0}{mm} in the direction of the thickness and width of the counter and between \SI{10}{mm} (at the counter ends) and \SI{100}{mm} in the direction of the length of the counter. 

Two additional lookup tables (for each counter length) are needed for Cerenkov photons in the scintillator and fibers. They use additional lookup parameters (the particle speed and particle direction) on which the energy, direction, and polarization of the Cerenkov photons depend. The number and energy distribution of the Cerenkov photons produced in each bin is based on the refractive indices of the material and the charge and speed of the particle.

\begin{figure}[H]
	\centering
	\includegraphics[width=0.6\textwidth,keepaspectratio]{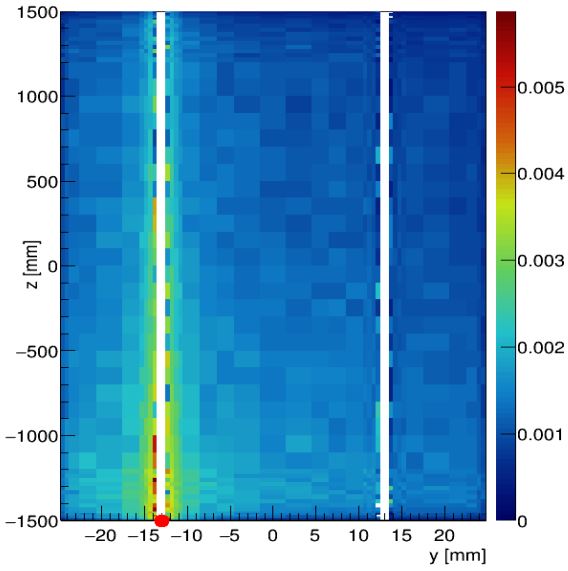}
	\caption{Example of a slice of the 3D grid of the detection probability lookup table for the SiPM marked with a red dot. Shown are the bins of a slice through the center of the CRV counter. The white stripes indicate the locations of the fiber channels which do not have scintillator material, and therefore cannot produce scintillation photons.}
	\label{fig:LookupTableDetectionProbabilities}
\end{figure}

The GEANT4 simulation transports the generated scintillator and Cerenkov photons through the scintillator and fibers all the way to the SiPMs (which have a wavelength-depending detection probability), and includes attenuation, reflection and wavelength-shifting processes. The probability of photons from each bin of the lookup tables to be detected by a SiPM is the fraction of the number of detected photons out of the number of simulated photons in each bin. These detection probabilities are stored in the lookup table. An example is shown in Fig.~\ref{fig:LookupTableDetectionProbabilities}.

The time a photon takes from its creation until reaching a SiPM has several contributions: the time of the scintillation process (if it is not a Cerenkov photon), the times of the wavelength-shifting processes in the scintillator and fibers, and the travel times inside the scintillator and fiber. These times --- excluding the time for the wavelength-shifting processes in the fibers --- are used to build a travel-time probability distribution for each lookup bin of the lookup table. The times required for the wavelength shifting processes (i.e., absorption and emission with a longer wavelength) in the fibers are treated separately. This reduces the time ranges required for the travel time probability distributions. Due to an overlap of the absorption and emission spectrum, there may be several absorption and re-emission processes for the same original photon. The number of emissions processes in the fibers for every generated photon is used to build an emission probability distribution for each lookup bin of the lookup table.

\section{CRV Hit Reconstruction}\label{ch:Reconstruction}

The hit reconstruction process (which is identical for simulated and real data) scans through the waveforms of each SiPM to find all local maxima. After subtracting the pedestal, the ADC values around these maxima are fitted with a function similar to the Gumbel (maximum) distribution~\cite{GumbelFunction}: 
\begin{equation}
ADC(t)=Ae^{-\frac{t-\mu}{\beta}-e^{-\frac{t-\mu}{\beta}}}.
\label{eq:Gumbel}
\end{equation}
An example of such a fit is shown in Fig.~\ref{fig:SampleEvent}. The time of the pulse peak is \(\mu\), the pulse height is \(A/e\), and the area under the curve is \(A\beta\). This area is proportional to the SiPM charge. This fact is used to convert the pulse area into the number of PEs. To find the conversion factor, a calibration needs to be done. 

\begin{figure}
	\centering
	\includegraphics[width=0.8\textwidth,keepaspectratio]{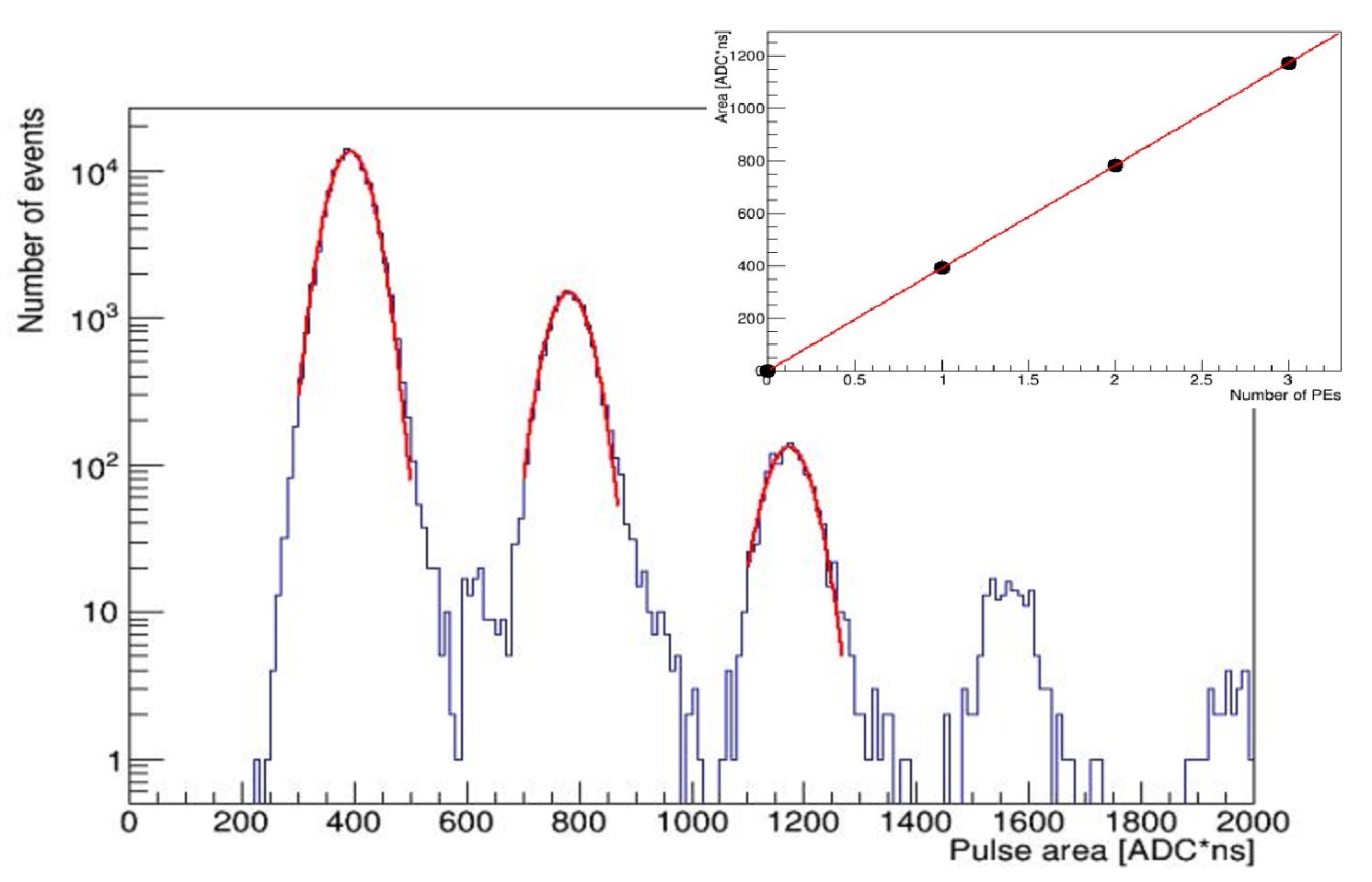}
	\caption{Reconstructed pulse areas of simulated dark noise (plots for real data look similarly). The fitted peaks of this distribution correspond to \SI{1}{PE}, \SI{2}{PE}, and \SI{3}{PE}. The relationship between pulse areas and number of PEs is shown in the inset plot, where a linear fit has been performed. The calibration constant is the inverse of the slope of the fitted function.}
	\label{fig:CalibrationHistogram}
\end{figure}

The calibration is done using dark-noise pulses. A dark-noise pulse of a fully charged pixel creates a SiPM charge that leads to a pulse whose area corresponds to \SI{1}{PE}. Occasionally, cross-talk may create simultaneous SiPM charges in more than one pixel. In these cases, the pulse area will correspond to \SI{2}{PE}, \SI{3}{PE}, or even more PEs. A histogram of these pulse areas shows peaks at the locations of \SI{1}{PE}, \SI{2}{PE}, \SI{3}{PE}, and so on (Fig.~\ref{fig:CalibrationHistogram}). Using these peak locations, the linear relationship between the pulse areas and the number of PEs can be extracted.

\section{Comparison with Test-Beam Data}\label{ch:ComparisonwithTestBeamData}

\begin{figure}
	\centering
	\includegraphics[width=0.7\textwidth,keepaspectratio]{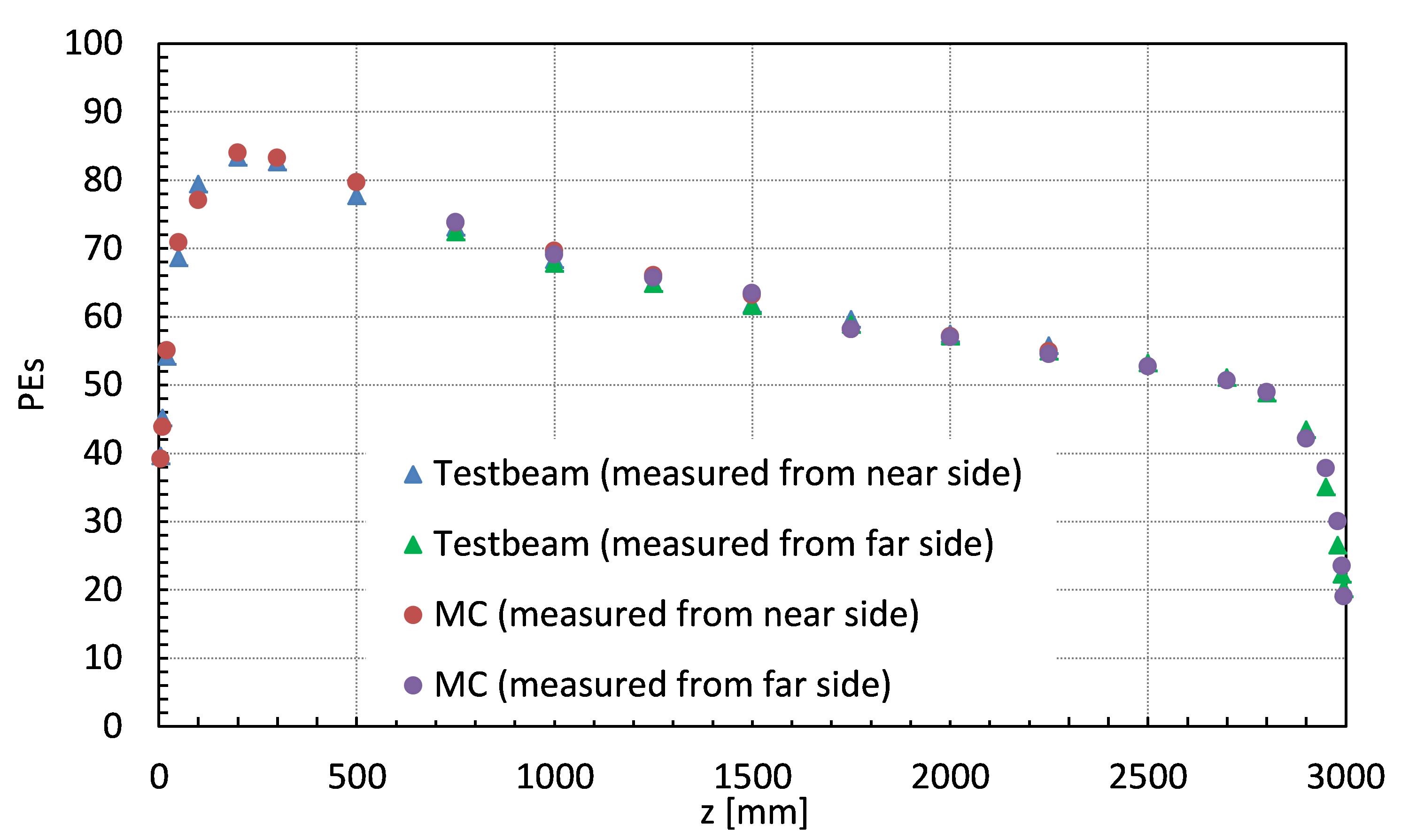}
	\caption{Comparison between simulation (using lookup tables) and test-beam data for the number of PEs for \SI{120}{GeV} protons normally incident at the center of a 3-m-long CRV counter and at different positions along a counter. The PE yields shown in the plot are averages of the PE yields of both SiPM at the same side of the counter. Due to the limited space at the test-beam area, only points between \SI{5}{mm} and \SI{2250}{mm} away from the SiPMs at the near side of the counter were targeted by the proton beam. However, the counters were also readout by SiPMs at the far side of the counter, for which these points are between \SI{2995}{mm} and \SI{750}{mm} away from the SiPMs. This made it possible to determine the PE yields for all points along the 3-m-long counter.}
	\label{fig:ComparisonPEsLookupTable}
\end{figure}

\begin{figure}
	\centering
	\includegraphics[width=0.7\textwidth,keepaspectratio]{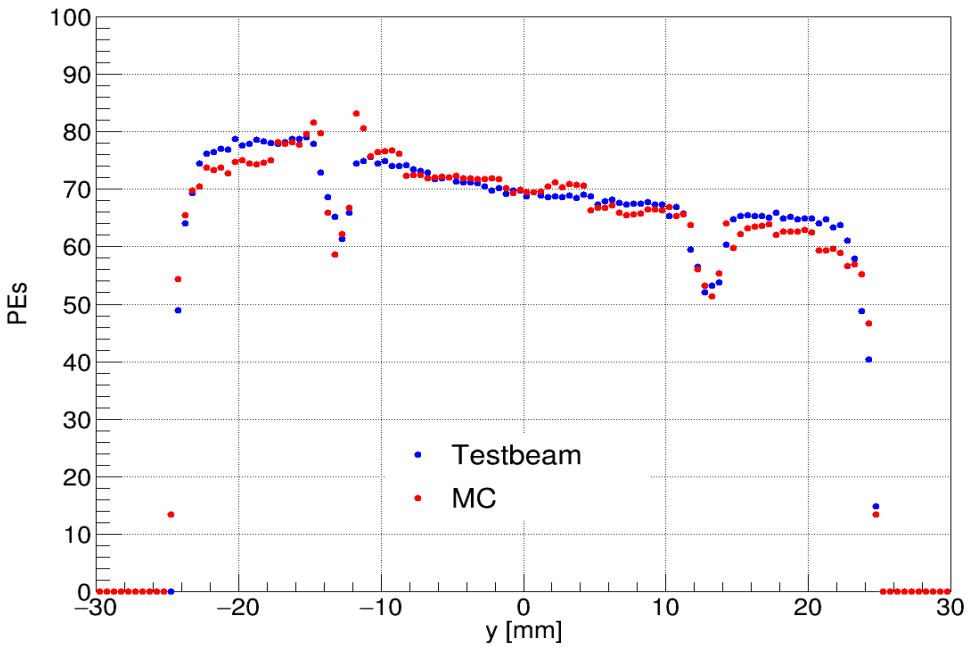}
	\caption{Comparison between simulation (using lookup tables) and test-beam data for the number of PEs for \SI{120}{GeV} protons normally incident at different locations across a 3-m-long CRV counter at a fixed longitudinal position (\SI{1}{m} away from the SiPM). The two dips indicate the locations of the fiber channels. In this plot, only the SiPM at the fiber at \SI{-13}{mm} is used. The discrepancies in the plot are not significant for our results.}
	\label{fig:ComparisonPEsTransverseScan}
\end{figure}

\begin{figure}
	\centering
	\includegraphics[width=0.7\textwidth,keepaspectratio]{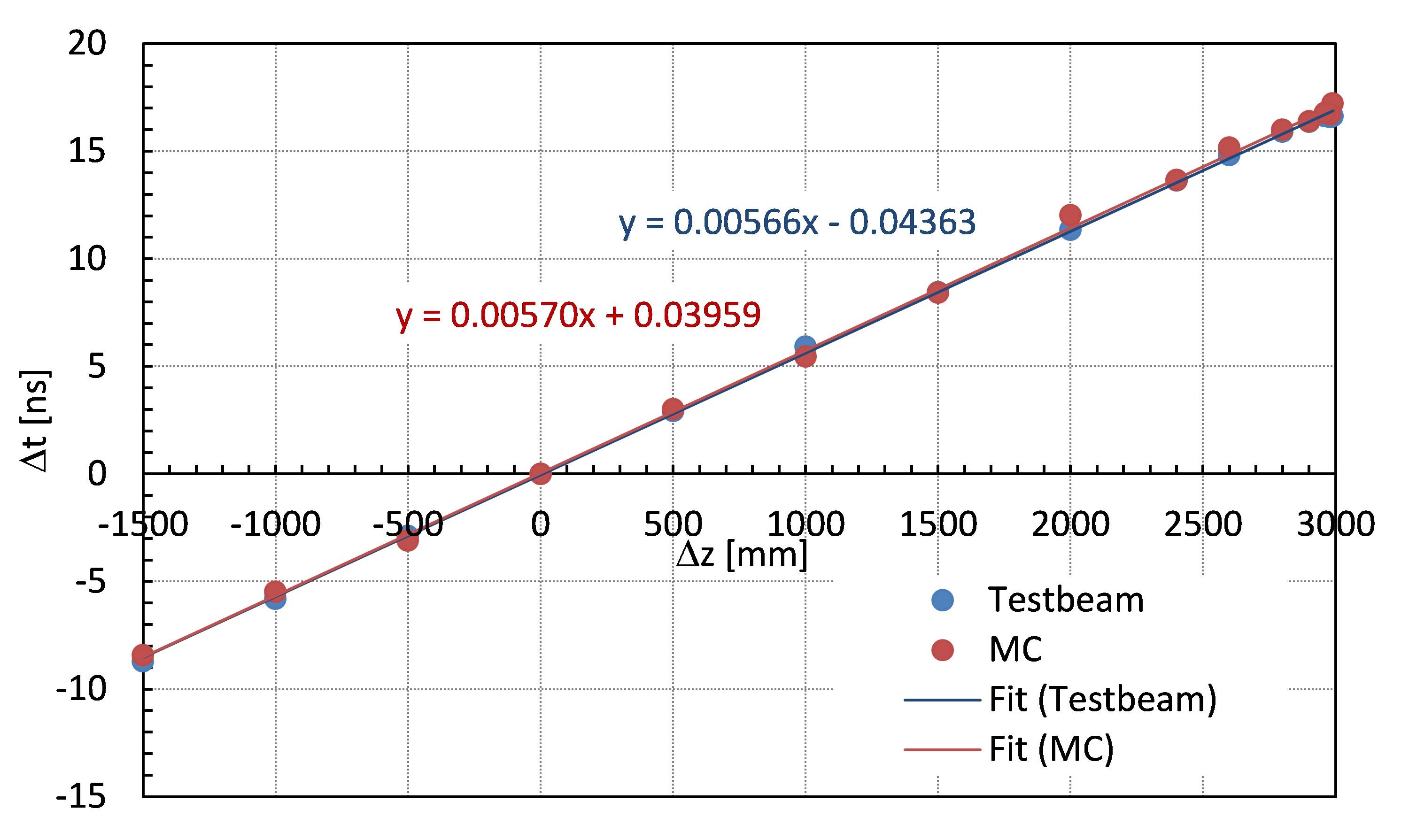}
	\caption{Comparison between simulation (using lookup tables) and test-beam data for the differences of the pulse times between both sides of a 3-m-long counter for \SI{120}{GeV} protons normally incident at different longitudinal locations along the counter. The horizontal axis shows the path length difference for the light traveling to both sides of the counter starting at the point where the proton hits the counter. The path length difference of 0 is at the center of the counter. The plot does not have any path length differences less than \SI{-1500}{mm}, because only points between \SI{5}{mm} and \SI{2250}{mm} away from the SiPMs at the near side of the counter (corresponding to \SI{2995}{mm} and \SI{750}{mm} away from the SiPMs at the far side) were measured due to the limited space at the test-beam area.}
	\label{fig:ComparisonPulseTimesLookupTable}
\end{figure}

The comparison between PE yields obtained using simulations with lookup tables and test-beam PE yields for \SI{120}{GeV} protons normally incident at a 3-m-long counter are shown in Fig.~\ref{fig:ComparisonPEsLookupTable} (at the center of the counter between the two fibers and at different positions along the counter) and Fig.~\ref{fig:ComparisonPEsTransverseScan} (at \SI{1}{m} away from the SiPM and at different positions across the counter). The agreement is remarkably good, even near the counter ends. The pulse time differences between both sides of the CRV counter were also compared (Fig.~\ref{fig:ComparisonPulseTimesLookupTable}). The slopes in this plot can be translated into speeds of light in the fiber of \SI{1.75e8}{m/s} (simulation) and \SI{1.77e8}{m/s} (test beam).

\section{Summary}\label{ch:Summary}

The simulation of the cosmic ray veto counters for the Mu2e experiment includes the simulation of scintillation and Cerenkov photons, attenuation, reflection, and wavelength-shifting processes. Lookup tables are used to reduce the computation time. The simulation of the SiPM response includes cross talk, dark noise, the recharge behavior of each pixel, and saturation effects. The output of the simulation is an ADC waveform. These waveforms are reconstructed to determine the time and the number of photoelectrons of each pulse. The reconstruction process is identical to what was used for test-beam studies. The comparison between the simulation and test-beam measurements shows that PE yields agree within 3\% (except at the counter ends), and that the simulated speed of light in the fiber differs only about 1\% from the measured value. 

\section*{Acknowledgments}

We are grateful for the valuable discussions with my Mu2e colleagues, and for the help from Craig Dukes, Craig Group, Sten Hansen, Yuri Oksuzian, Anna Pla-Dalmau, Paul Rubinov, Enhao Song, and Yongyi Wu. I thank Mandy Rominsky, the crew of the Fermilab Test Beam Facility, and everyone who took shifts at the June 2017 test beam. We are grateful for the vital contributions of the Fermilab Computing Division, the Fermilab staff and the technical staff of the participating institutions. This work was supported by the US Department of Energy, the Italian Istituto Nazionale di Fisica Nucleare, the US National Science Foundation, the Ministry of Education and Science of the Russian Federation, the Thousand Talents Plan of the Republic of China, the Helmholtz Association of Germany, and the EU Horizon 2020 Research and Innovation Program under the Marie Sklodowska-Curie Grant Agreement No.690385. Fermilab is operated 
by Fermi Research Alliance, LLC under Contract No. DE-AC02-07CH11359 with the US Department of Energy.

\begingroup
\raggedright
\bibliography{DPF2019_bibfile}
\endgroup

\end{document}